\documentclass[aps,twocolumn,pre,floatfix]{revtex4-2}

\usepackage{booktabs}

\usepackage{xcolor,hyperref}
\hypersetup{
   colorlinks,
   linkcolor={blue!50!black},
   citecolor={blue!50!black},
   urlcolor={blue!80!black}
}

\usepackage{epsfig, amsmath}
\usepackage{bm}
\usepackage{color}
\usepackage{soul,ulem}
\usepackage{hyperref}
\usepackage{footmisc}
\usepackage{wrapfig2,graphicx,enumitem}
\DeclareMathAlphabet{\mathitbf}{OML}{cmm}{b}{it}

\newcommand{\zerovector}{\bm{0}}
\newcommand{\dv}{\mathitbf d}

\newcommand{\uv}{\mathitbf u}

\newcommand{\xv}{\mathitbf x}

\newcommand{\zv}{\mathitbf z}

\newcommand{\piv}{\bm{\pi}}
\newcommand{\psiv}{\bm{\psi}}
\newcommand{\Psiv}{\bm{\Psi}}

\newcommand{\calBold}[1]{\mbox{\boldmath${\cal #1}$}}
\newcommand{\dbar}{{\,\mathchar'26\mkern-12mu d}}

\begin{document}

\title{Effect of molecular constraints on vibrational and quasilocalized excitations in glasses}
\author{Keane Ramdin}
\author{Edan Lerner}
\email{e.lerner@uva.nl}

\affiliation{Institute for Theoretical Physics, University of Amsterdam, Science Park 904, 1098 XH Amsterdam, the Netherlands}

\begin{abstract}
Recent years have seen substantial progress in elucidating the statistical physics of the vibrational properties of structural glasses. Although many real-world glasses relevant for science and technology are molecular, the majority of computational studies concerning the mechanical and vibrational properties of structural glasses employ simple atomistic glass-forming models. Thus, the effects of stiff molecular constraints on mechanical and vibrational glass physics remain largely unexplored. In this work, we directly compare the properties of a molecular computer glass with those of an atomistic model featuring the same inter-molecular interaction potential, and created using the same formation protocol. We find that the molecular glass features a higher degree of mechanical disorder, with larger mesoscopic correlation lengths, while at the same time it hosts a lower number of soft, quasilocalized vibrations per atom -- compared to the atomistic glass. We rationalize these differences by accounting for the reduction in the effective number of degrees of freedom induced by the stiff molecular constraints. We additionally find that nonlinear plastic modes --- that carry plastic deformation in driven glassy solids --- couple much more strongly to volumetric strains in the molecular glass. Several future research directions are discussed. 
  
\end{abstract}

\maketitle

\section{i\lowercase{ntroduction}}
\vspace{-0.2cm}

Structural glasses formed by cooling a melt generically feature both positional/structural disorder, and mechanical frustration, which together give rise to rich vibrational spectra that have been puzzling workers in the field for decades~\cite{soft_potential_model_1987,ohern2003,parisi_boson_peak_2003,matthieu_thesis,Schirmacher_prl_2007,tanaka_boson_peak_2008,Chumakov_2011_bosonPeak,LB_modes_2019,JCP_Perspective,Beltukov_prb_2021,nonlinear_dispersion_arxiv_2026}. Due to the translational symmetry of these materials' Hamiltonians, at low frequencies these systems feature long-wavelength phononic excitations -- dressed by noise due to the underlying structural disorder and mechanical frustration~\cite{Schirmacher_prl_2007,phonon_widths,Beltukov_prb_2021}. These phononic excitations follow a nonlinear dispersion curve whose form is sensitive to the degree of underlying mechanical disorder, which, in turn, gives rise to an excess of vibrational excitations on top of Debye's phononic spectrum~\cite{Schirmacher_prl_2007,Beltukov_prb_2021,nonlinear_dispersion_arxiv_2026}.

In addition, glassy solids quenched from a melt generically host a population of soft, quasilocalized vibrations (QLVs) at low frequencies, which abide by a universal nonphononic spectrum that grows from zero frequency $\omega$ as $\sim\!\omega^4$~\cite{soft_potential_model_1987,Gurevich2003,JCP_Perspective}. These excitations were shown to control wave attenuation rates~\cite{grzegorz_damping_defects_2025}, plastic deformation under external loading~\cite{micromechanics2016}, and argued to explain glasses' cryogenic specific heat~\cite{soft_potential_model_1991}. Their number density at low frequencies was shown to depend dramatically on the thermal history of the ancestral liquid from which their host glass was formed~\cite{pinching_pnas,LB_modes_2019}. Recent work~\cite{boson_peak_2d_jcp_2023,nonlinear_dispersion_arxiv_2026} established that both phononic and nonphononic excitations populate the so-called boson peak that is universally observed in the vibrational spectrum of glassy solids~\cite{ramos2023low}. 



Many laboratory~\cite{ultrastable_review_2007,Rodriguez-Tinoco2022}, technologically relevant~\cite{technological1,technological2} and industrially relevant~\cite{industrial1,industrial2} glasses are molecular, and the anomalous vibrational properties of such glasses have been documented experimentally for decades. Inelastic neutron and Raman scattering studies have characterized the boson peak in molecular and polymeric glass formers~\cite{Niss2007,Hong2008}, and the excess low-frequency modes of vitreous silica were shown early on to involve coupled rotations of rigid SiO$_4$ tetrahedra~\cite{Buchenau1984,Buchenau1986} --- an indication that the internal rigidity of molecular units plays a role in shaping the low-frequency spectrum. More recently, low-temperature specific-heat measurements in ultrastable organic glasses have revealed a strong sensitivity of low-energy excitations to the glass's thermal history~\cite{PerezCastaneda2014}, echoing the dependence of quasilocalized modes on glass stability found in simple computer glasses~\cite{LB_modes_2019,pinching_pnas}.

On the computational side, molecular glass formers have been studied extensively in the context of supercooled-liquid dynamics --- the rigid three-site model of ortho-terphenyl of Lewis and Wahnstr\"om~\cite{Lewis1994} being a prominent example --- and the vibrational properties of model polymer glasses have received growing attention~\cite{Giuntoli2018,Tomoshige2019,Tomoshige2021}, including the observation that the boson peak of a polymer glass is governed by non-bonding interactions alone~\cite{Giuntoli2018}, and that soft, quasilocalized modes anticipate the loci of structural rearrangements in polymer glasses~\cite{Smessaert2014,Smessaert2015}. The universal $\sim\!\omega^4$ nonphononic spectrum has also been shown to emerge in triatomic molecular and polymeric models~\cite{modes_prl_2020}. Nonetheless, the majority of computational studies regarding vibrational physics and plasticity in glassy solids employ simple, atomistic glass forming models (see, however, e.g.~Refs.~\cite{tanaka_boson_peak_2008,Tanguy_2016}), and a controlled, equal-footing comparison between a molecular glass and an atomistic glass --- one that isolates the effect of stiff intramolecular constraints from differences in interaction potential, density, or formation history --- has, to the best of our knowledge, not been carried out. This leaves open the question: what differences in the vibrational and elastoplastic properties of a glass are observed once stiff, intramolecular interactions are at play?

In this work we address this question by constructing and studying a model molecular glass former, and a model atomistic glass former, both employing the same intermolecular interaction potential. Furthermore, we create both ensembles (atomistic and molecular) of glasses by quenching high-energy states, in a regime in which the properties of the resulting glasses lose all dependence on the ancestral, high-energy liquid states~\cite{boring_paper,cge2_jcp2020}. Finally, we tune the density of the ancestral liquids of the two models such that the resulting glasses created by an instantaneous quench feature a vanishing ratio of hydrostatic pressure to bulk modulus. These choices are made in order to allow an equal-footing comparison of the elastic and vibrational properties between the atomistic and molecular glass formers, and to single out --- in the cleanest possible way --- the effect of molecular constraints on glassy vibrational physics. 

We find that introducing stiff molecular constraints leaves the universal, well-known features of glassy vibrational spectra essentially intact, while systematically amplifying the degree of mechanical disorder in the glass. The molecular glass is elastically stiffer than the atomistic one, yet at the same time more mechanically disordered. Most notably, we find that the excitations responsible for plastic deformation in glasses~\cite{micromechanics2016,episode_1_2020} couple far more strongly to volume-changing (dilatational) deformation in the molecular glass than in the atomistic one — suggesting a possible role for molecular constraints in dilatational plasticity.

This paper is structured as follows. In Sect.~\ref{sec:models_and_methods} we present our molecular and atomistic glass forming models, and review the units in which various observables are reported in this work. In Sect.~\ref{sec:elasticity} we review the macroscopic elastic properties of the models. Sect.~\ref{sec:vibrations} presents a comparative study of various vibrational properties of the models. Sect.~\ref{sec:nonlinear_modes} compares the properties of nonlinear plastic modes between the two models. A summary, discussion, and future research questions are provided in Sect.~\ref{sec:discussion}.

\section{M\lowercase{odels, protocols and units}}
\label{sec:models_and_methods}
\vspace{-0.2cm}

\subsection{Molecular glass}
\vspace{-0.2cm}

We employ a molecular glass forming model inspired by Ref.~\cite{berthier_molecular_glass}. We place $N_{\rm m}\!=\!N/3$ molecules in a periodic simulation box of volume $V$, each comprising three atoms of equal mass $m$ such that the total number of atoms is $N$. All atoms interact via a conventional Lennard-Jones (LJ) potential that reads
\begin{equation}
    \varphi_{\mbox{\tiny LJ}}(r) = 4\varepsilon\left(\frac{1}{\tilde{r}^{12}} - \frac{1}{\tilde{r}^6} + c_6\tilde{r}^6 + c_4\tilde{r}^4 + c_2\tilde{r}^2 + c_0\right)\,,
\end{equation}
where $r$ is the pairwise interatomic separation, $\tilde{r}\!\equiv\! r/\sigma$ is a dimensionless interatomic separation, $\sigma$ is a microscopic length, and $\varepsilon$ is an energy scale. The dimensionless coefficients $c_{2\ell}$ guarantee that the potential tapers down to zero smoothly at $r_{\rm cutoff}/\sigma\!=\!2.0$. The coefficients are provided in the table below. 
\begin{center}
\begin{tabular}{c r}
\hline
coefficient & numerical value \\
\hline
$c_{0}$ & $0.2919921875$ \\
$c_{2}$ & $-0.16259765625$ \\
$c_{4}$ & $0.0322723388671875$ \\
$c_{6}$ & $-0.002227783203125$ \\
\hline
\end{tabular}
\end{center}

\begin{figure}[ht!]
  \includegraphics[width = 0.4\textwidth]{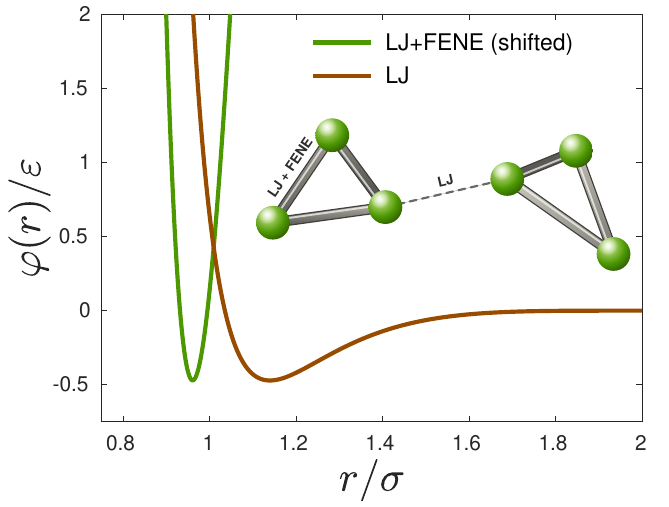}
  \caption{\footnotesize The intermolecular (LJ) and intramolecular (LJ + FENE) pairwise potentials employed in our molecular model. Our atomistic binary glass model employs the LJ potential but two effective particle sizes, see text for details.}
  \label{fig:potential_fig}
\end{figure}

In addition to the LJ interaction, atoms within the same molecule interact via the FENE potential~\cite{FENE}
\begin{equation}
    \varphi_{\mbox{\tiny FENE}}(r) = -\frac{1}{2}kR_0^2\ln(1-r^2/R_0^2)\,.
\end{equation}
We choose $k\!=\!40\varepsilon/\sigma^2$ and $R_0\!=\!r_{\mbox{\tiny cutoff}}$. With these parameters, we find that the ratio of the stiffness $\varphi''$ evaluated at the minimum of the LJ potential and at the minimum of the LJ+FENE potentials is about 1:30. The employed intra- and inter-molecular potentials are plotted in Fig.~\ref{fig:potential_fig}. 

To create molecular glasses, we position our molecules on a fcc lattice, and launch NVE dynamics for 10.0$\sqrt{m\sigma^2/\varepsilon}$, followed by an instantaneous quench to zero temperature. We set the number density $N/V\!=\!1.05\sigma^{-3}$ such that the resulting pressure-to-bulk-modulus ratio of our molecular glasses is vanishingly small. Most of the analysis presented in what follows is applied to an ensemble of 10,000 glasses of $N_{\rm m}\!=\!1372$ molecules, while larger samples of $N_{\rm m}\!=\!23328\!=\!69984/3$ molecules were also created and analyzed (cf.~Sect.~\ref{sec:dispersion}).

\subsection{Atomistic glass}
\vspace{-0.2cm}

In order to distill the effects of stiff, molecular constraints, we also employ an atomistic, 50:50 binary glass forming model in which `large' and `small' particles interact with the same intermolecular LJ potential described above. We employ weakly non-additive interaction length coefficients of $\sigma_{\mbox{\tiny LL}}\!=\!1.4\sigma$, $\sigma_{\mbox{\tiny LS}}\!=\!1.18\sigma$, and $\sigma_{\mbox{\tiny SS}}\!=\!1.0\sigma$, where the subscripts `LL',`LS' and `SS' denote large-large, large-small and small-small interactions, respectively. The mass of all particles is $m$. 

To create atomistic glasses, we place `large' and `small' atoms randomly on a fcc lattice, run NVE dynamics for 10.0$\sqrt{m\sigma^2/\varepsilon}$, followed by an instantaneous quench to zero temperature.
We set the number density at $N/V\!=\!0.535\sigma^{-3}$, such that the resulting pressure-to-bulk-modulus ratio of our atomistic glasses is vanishingly small. Most of the analysis presented in what follows is applied to glasses of $N\!=\!1372$ atoms, while larger samples of $N\!=\!70304$ atoms were also created and analyzed (cf.~Sect.~\ref{sec:dispersion}).

\subsection{Units}
\label{sec:units}
\vspace{-0.2cm}
In order to facilitate an equal-footing comparison in what follows between our molecular and atomistic glasses, for both models we report lengths in terms of $a_0\!\equiv\!(V/N)^{1/3}$ (and not $\sigma$ used above), where $V$ denotes the volume and $N$ the number of atoms. Additionally, frequencies are expressed in terms of $\sqrt{\varepsilon/ma_0^2}$ and elastic moduli in terms of $\varepsilon/a_0^3$.

\subsection{Differences between the two models}
\label{sec:model_differences}
\vspace{-0.2cm}
Before proceeding, we stress that the two models differ in respects other than molecular connectivity alone. First, the atomistic model requires a binary size disparity in order to frustrate crystallization, whereas the molecular model consists of identical atoms whose glass-forming ability stems from connectivity. The two models therefore do not share the same structural disorder at the particle level. Second, the number densities of the two models differ, since they are tuned to yield a vanishing pressure-to-bulk-modulus ratio rather than matched directly. Third, in the molecular model the intramolecular bond length nearly coincides with the intermolecular nearest-neighbor distance (cf.~Fig.~\ref{fig:potential_fig}), a feature that has no counterpart in the atomistic model.

A perfectly controlled comparison --- in which molecular constraints are the \emph{only} difference between the two systems --- is not attainable: a monodisperse atomistic Lennard-Jones system crystallizes under our formation protocol, while introducing size disparity into the molecular model would add a second source of disorder. Our choices --- a shared intermolecular potential, a shared formation protocol, matched vanishing $p/K$, and lengths reported in units of $a_0$ --- are designed to minimize the remaining differences. Nonetheless, when we attribute differences between the two glasses to molecular constraints in what follows, this should be understood as the most plausible interpretation rather than a strictly established causal link.

\section{M\lowercase{acroscopic elastic properties}}
\label{sec:elasticity}
\vspace{-0.2cm}

Before presenting our comprehensive comparative study of the vibrational properties of our two models, we first present their macroscopic elastic properties. In Table~\ref{table:networks} we report the ensemble average of the shear modulus $G$, the bulk modulus $K$, and the Poisson's ratio $\nu\!\equiv\!(3K\!-\!2G)/(6K+2G)$. Microscopic expressions for athermal elastic moduli can be found, e.g., in~\cite{boring_paper}. We also report the rescaled standard deviation of the shear modulus, denoted $\chi$, namely
\begin{equation}\label{eq:chi}
    \chi \equiv \sqrt{N}\frac{\Delta G}{\langle G \rangle}\,,
\end{equation}
where $\Delta G$ and $\langle G \rangle$ denote the sample-to-sample standard deviation and mean of the shear modulus, respectively. $\chi$ is a broadly applicable quantifier of mechanical disorder~\cite{karina_chi_paper_2023,julia_chi_2024}, related to spectral widths~\cite{jcp_letter_scattering_2021,grzegorz_2025_scattering_perspective} and dispersion~\cite{Schirmacher_prl_2007,Beltukov_prb_2021,nonlinear_dispersion_arxiv_2026} of elastic waves. It typically assumes values between 1-4 in conventional atomistic computer glasses~\cite{karina_chi_paper_2023}. Here $\chi$ is calculated over our ensembles of 10,000 glass samples of each model, using the outlier exclusion approach detailed in Ref.~\cite{sticky_spheres1_karina_pre2021}.

\begin{table}[h]
    \centering
    \caption{Macroscopic elastic properties}
    \label{table:networks}
    \begin{tabular}{| c | c c c c|}
        \hline
        model & $G$ & $K$ &$\nu$ & $\chi$ \\ \hline
        molecular & 12.9 & 66.3 & 0.41 & 5.1 \\ \hline
        atomistic & 10.3 & 48.9 & 0.40 & 4.0  \\ \hline
    \end{tabular}
\end{table}

A few interesting observations emerge. First, we find that both the shear and bulk moduli of the molecular glass are larger than those of the atomistic glass. Naively, one might expect this due to the presence of stiff, intramolecular interactions; however, the larger degree of mechanical disorder --- as quantified via $\chi$ --- could lead to an opposite expectation. 

Surprisingly, we find the ratio between the molecular to atomistic average shear moduli, $G_{\mbox{\tiny mol}}/G_{\mbox{\tiny atom}}\!\approx\!1.25$, to be very slightly smaller than the bulk moduli ratio $K_{\mbox{\tiny mol}}/K_{\mbox{\tiny atom}}\!\approx\!1.35$. Here too, the opposite could be expected since the stiff, molecular constraints are expected to lead to enhanced nonaffinity under compression~\cite{compression_nonaffinity_pre_2019} in the molecular glass, which, in turn, would reduce (in relative terms) the bulk modulus of the molecular glass compared to the atomistic one. As a result of the aforementioned relatively larger bulk modulus of the molecular glass, its Poisson's ratio is very slightly higher than the atomistic glasses' Poisson's ratio.

Finally, we note that the quantifier $\chi$ of mechanical disorder is larger in the molecular glass. Since $\chi$ is understood to represent the square-root of a mesoscopic correlation volume~\cite{julia_chi_2024,nonlinear_dispersion_arxiv_2026} (and see further discussion in Sect.~\ref{sec:dispersion}), and we expect this volume to be larger in the molecular glass due to the presence of stiff intramolecular interactions, the larger $\chi$ of the molecular glasses is in accordance with our expectations.

\section{V\lowercase{ibrational properties}}
\label{sec:vibrations}
\vspace{-0.2cm}

\subsection{Full vibrational density of states}
\label{sec:full_vdos}
\vspace{-0.2cm}

\begin{figure}[ht!]
  \includegraphics[width = 0.5\textwidth]{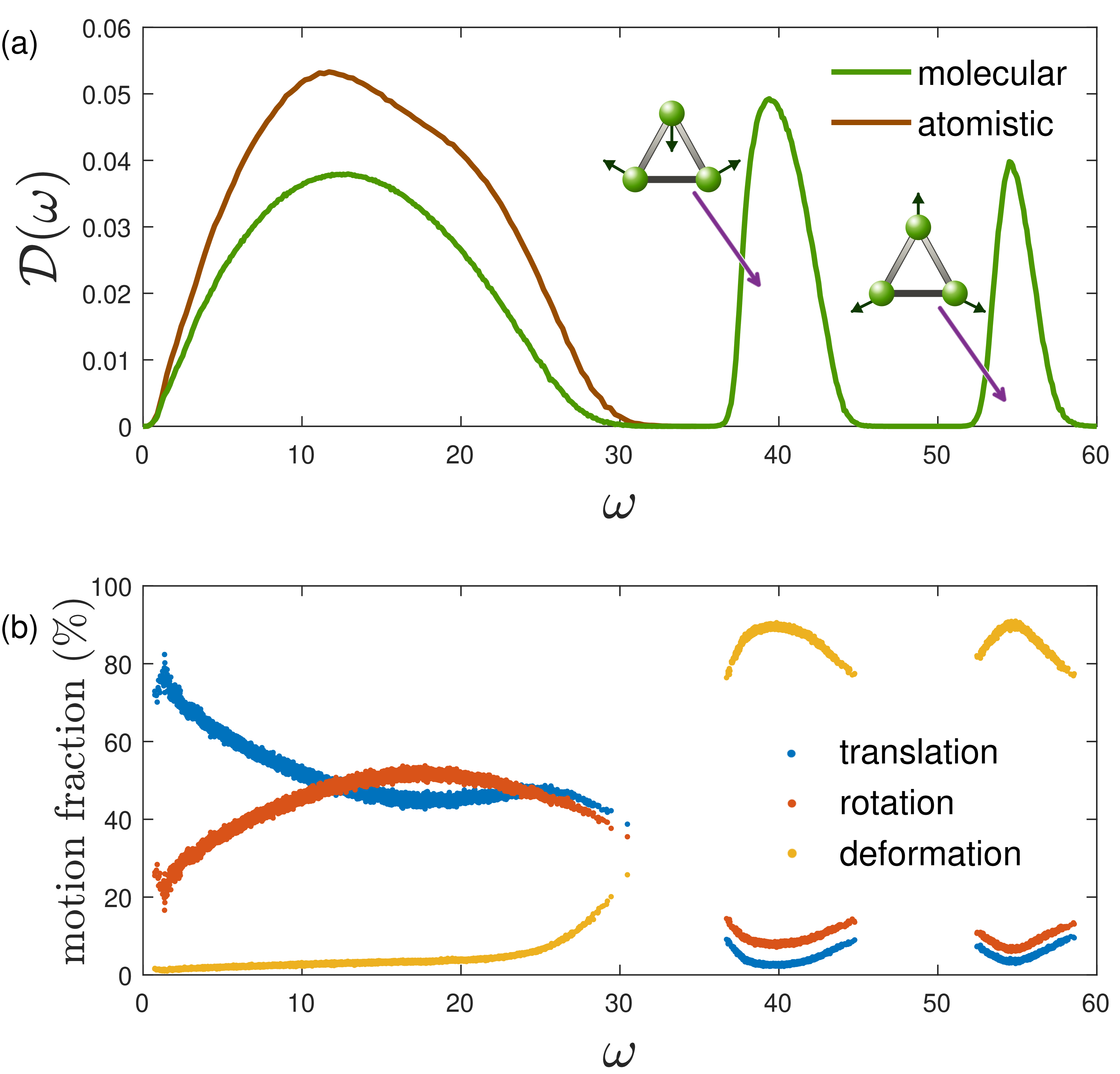}
  \caption{\footnotesize (a) Vibrational density of states (VDoS) ${\cal D}(\omega)$ for the molecular and atomistic glasses, see figure legend. (b) Breakdown of the molecular motion of vibrational modes into translational, rotational and intramolecular deformation, per vibrational mode of frequency $\omega$, expressed in percentages, and see Appendix~\ref{sec:molecular_motion_appendix} for definitions.}
  \label{fig:full_vdos}
\end{figure}

We kick off the discussion by presenting the full vibrational density of states (VDoS) ${\cal D}(\omega)$ of the two glasses in Fig.~\ref{fig:full_vdos}a. In Fig.~\ref{fig:full_vdos}b we decompose the motion encoded in the vibrational modes of the molecular glass into molecular translation, rotation and deformation, expressed as percentages and plotted against frequency $\omega$; definitions and details are provided in Appendix~\ref{sec:molecular_motion_appendix}. Our data for the VDoS were calculated over an ensemble of 100 independent glass realizations of each of our models, while the motion decomposition shown in Fig.~\ref{fig:full_vdos}b was computed on a single molecular glass realization.

These results validate two key expectations: first, we find that the molecular VDoS features a trimodal structure. The main component at low frequencies corresponds to vibrations in which the molecules translate and rotate, but do not deform --- reminiscent of the classical picture of the low-frequency vibrations of vitreous silica as coupled librations of nearly rigid SiO$_4$ tetrahedra~\cite{Buchenau1984,Buchenau1986,Taraskin1997}. The two higher-frequency bands of the molecular $\mathcal{D}(\omega)$
correspond to the intramolecular deformation modes of our three-atom
molecules --- sometimes called `optical modes'~\cite{optical_modes_prb_1977,Taraskin1997epl}. Each molecule possesses three internal modes: a doubly degenerate bending
mode, which gives rise to the band at $\omega\!\approx\!40$, and a
stiffer breathing mode, which gives rise to the band at
$\omega\!\approx\!55$; both are illustrated in the cartoon insets of
Fig.~\ref{fig:full_vdos}a. Consistent with this assignment, the lower of the
two bands carries approximately twice the spectral weight of the higher
one, reflecting the two-fold degeneracy of the bending mode. These results validate that our choice of stiffness $k$ of the FENE potential leads to a clean scale separation between high-frequency, intramolecular vibrations, and those involving the translation and rotation of molecules, in accordance with our aim.

Second, the very good overlap in frequency range between the low-frequency band in the molecular model, and the total VDoS of the atomistic model,  lends support to the meaningfulness of our equal-footing comparison between the two models. This is achieved by employing the same intermolecular LJ potential in the atomistic glass, by setting the density such that the hydrostatic pressure of the two glasses is vanishingly small, by employing the same glass-formation protocol for the two models, and by expressing frequencies in terms of $\sqrt{\varepsilon/ma_0^2}$ as explained in Sect.~\ref{sec:units}.

\subsection{Wave dispersions}
\label{sec:dispersion}
\vspace{-0.2cm}

\begin{figure}[ht!]
  \includegraphics[width = 0.5\textwidth]{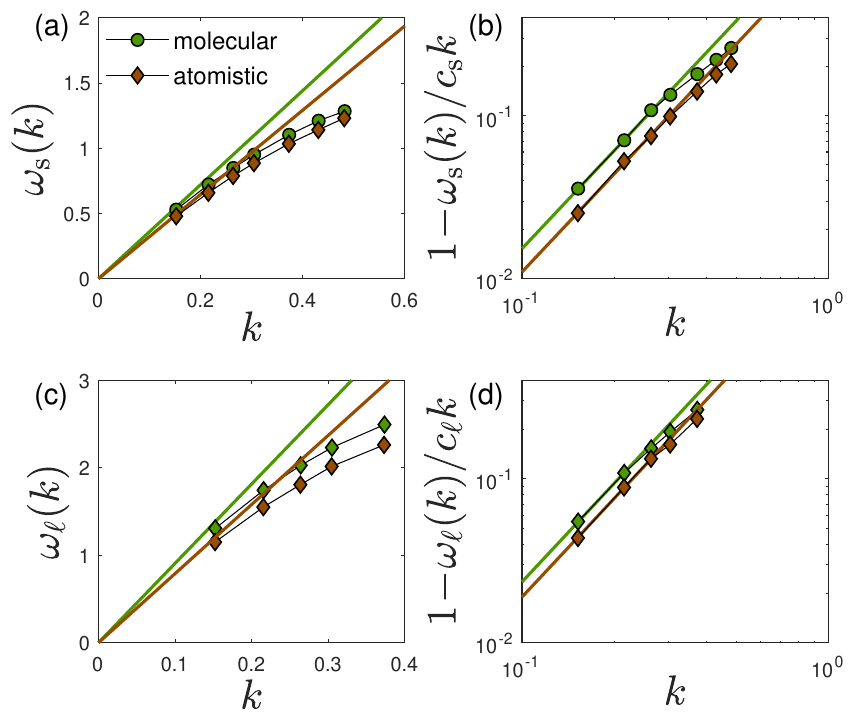}
  \caption{\footnotesize (a) Dispersion $\omega_{\rm s}(k)$ of transverse waves for both glass models as detailed in the figure legend. (b) Following Eq.~(\ref{eq:dispersion}) we plot $1\!-\!\omega_{\rm s}(k)/c_{\rm s}k\!\sim\!\xi_{\rm s}^2k^2$ at low $k$, see text for discussion.
  Panels (c),(d) are the same as (a),(b) but for longitudinal waves.}
  \label{fig:dispersion}
\end{figure}

We next analyze and compare the wave dispersion behavior of the two glass models. To this aim, we employ the `imposed-wave method' as described in Ref.~\cite{IWM_paper}; the method measures the effective frequency $\omega_{{\rm s},\ell}(k)$ of transverse (`s') or longitudinal (`$\ell$') wave-like excitations, given some wavenumber $k$. For these measurements, we employed systems of $N_{\rm m}\!=\!23328$ molecules and $N\!=\!70304$ atoms for the molecular and atomistic glasses, respectively. The results are presented in Fig.~\ref{fig:dispersion}a for transverse waves, and in Fig.~\ref{fig:dispersion}c for longitudinal waves.

The elastic wave dispersion curves follow a form~\cite{nonlinear_dispersion_arxiv_2026}
\begin{equation}\label{eq:dispersion}
    \omega_{{\rm s},\ell}(k) \simeq c_{{\rm s},\ell}k - c_{{\rm s},\ell}\Upsilon\xi_{{\rm s},\ell}^2k^3\,,
\end{equation}
where $k$ is the wavenumber, $\Upsilon\!=\!1/96$ is a dimensionless prefactor inspired by the nonlinear wave dispersion of a 1D chain, and $\xi_{{\rm s},\ell}$ is a lengthscale that characterizes the degree of mechanical disorder of the glass~\cite{nonlinear_dispersion_arxiv_2026}. Therefore, by tuning $c_{{\rm s},\ell}$ and plotting $1\!-\!\omega_{{\rm s},\ell}/c_{{\rm s},\ell}k\!\approx\!\Upsilon\xi_{{\rm s},\ell}^2k^2$ on logarithmic scales in Figs~\ref{fig:dispersion}b,d, we are able to extract estimates for $c_{{\rm s},\ell}$ and $\xi_{{\rm s},\ell}$, which are reported in Table~\ref{table:waves}.

\begin{table}[h]
    \centering
    \caption{Wave dispersion properties}
    \label{table:waves}
    \begin{tabular}{| c | c c c c |}
        \hline
        model & $c_{\rm s}$ & $c_{\rm \ell}$ & $\xi_{\rm s}$ & $\xi_{\rm \ell}$    \\ \hline
        molecular & 3.60 & 9.07 & 12.1 & 15.0  \\ \hline
        atomistic & 3.23 & 7.90 & 10.3 & 13.5  \\ \hline
    \end{tabular}
\end{table}

We first compare the extracted wave-speeds from the wave dispersions with those obtained from the elastic moduli; the latter yield $c_{\rm s}\!=\!\sqrt{G/\rho}\!\approx\!3.6$ ($\rho$ denotes the mass density) and $c_\ell\!=\!\sqrt{(K+4G/3)/\rho}\!\approx\!9.1$ for the molecular glass, and $c_{\rm s}\!\approx\!3.2$ and $c_\ell\!\approx\!7.9$ for the atomistic glass, in very good agreement with the values extracted from the dispersion reported in Table~\ref{table:waves}. 

We next note that $\xi_{{\rm s,mol}}/\xi_{{\rm s,atom}}\!\approx\!1.17$, consistent with our finding that $\chi$ is larger in the molecular glass compared to the atomistic glass. Moreover, $\xi_{\rm s}$ is expected to scale as $a_0\chi$~\cite{nonlinear_dispersion_arxiv_2026,footnote2,breakdown}; this ratio is very roughly consistent with $\chi_{\rm mol}/\chi_{\rm atom}\!\approx\!1.27$. $\chi$ is hard to extract reliably (see discussion in Ref.~\cite{phonon_widths2}), which might explain the discrepancy between the two ratios.

\subsection{Soft quasilocalized vibrations}
\label{sec:qlv}
\vspace{-0.2cm}

Recent work (see~\cite{JCP_Perspective} for a comprehensive review) has established that the low-frequency nonphononic spectra of glasses, disordered crystals~\cite{disordered_crystals_prl_2022}, and other disordered solids~\cite{frustrated_networks_pre_2024} follow a universal quartic law, of the form ${\cal D}_{\rm g}(\omega)\!\simeq A_{\rm g}\omega^4$, where $A_{\rm g}$ is a nonuniversal prefactor that depends on the state of disorder of the solid. The excitations that populate this nonphononic tail are quasilocalized; they feature a disordered core of linear size $\xi_{\rm g}$, decorated by algebraically decaying fields $\sim\!r^{1-\dbar}$ at distance $r$ away from the core, in $\dbar$ spatial dimensions~\cite{modes_prl_2016}. These algebraic decays for $\dbar\!\ge\!3$ lead to a $\sim\!1/N$ scaling of the participation ratio $e$~\cite{footnote} of these modes (with logarithmic corrections in 2D), indicating their (quasi)localization. Thus, the product $Ne$ is a measure of the spatial extent of quasilocalized vibrations.

The nonuniversal prefactor $A_{\rm g}$ of the nonphononic VDoS ${\cal D}_{\rm g}(\omega)$ has dimensions of [frequency]$^{-5}$, and can be written as $A_{\rm g}\!=\!{\cal N}\omega_{\rm g}^{-5}$~\cite{cge_paper,pinching_pnas}, where $\omega_{\rm g}$ is a characteristic frequency scale of quasilocalized vibrations, and ${\cal N}$ is proportional to their number per degree of freedom. In~\cite{cge_paper} it was proposed and established that the stiffness associated with the linear displacement response to a local force dipole is a good proxy for $\omega_{\rm g}^2$. Using these results, it was shown in~\cite{pinching_pnas} that ${\cal N}$ follows a Boltzmann-like scaling with the parent equilibrium temperature $T_{\rm p}$ from which glassy solids were instantaneously quenched.

\begin{figure}[ht!]
  \includegraphics[width = 0.5\textwidth]{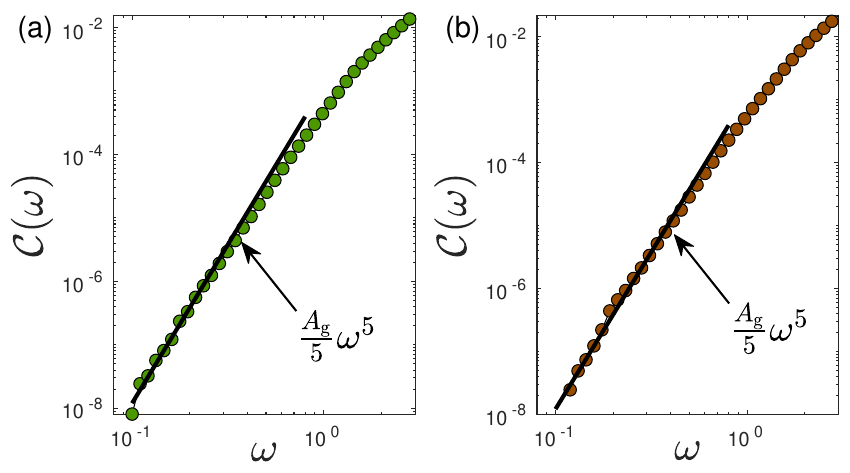}
  \caption{\footnotesize The cumulative VDoS ${\cal C}(\omega)\!=\!\int_0^\omega{\cal D}(\omega')d\omega'$ for our (a) molecular glass and (b) atomistic glass, see text for discussion.}
  \label{fig:Ag_fig}
\end{figure}

Here we aim at comparing all of the aforementioned properties and observables related to the statistical physics of quasilocalized vibrational modes -- between our molecular and atomistic glasses. To this aim, we employ our ensembles of 10,000 glasses of relatively small sizes, of both models. This choice allows us to push phononic excitations to higher frequencies, and cleanly expose the nonphononic VDoS~\cite{modes_prl_2016}. We first plot in Fig.~\ref{fig:Ag_fig} the cumulative VDoS, defined as ${\cal C}(\omega)\!=\!\int_0^\omega{\cal D}(\omega')d\omega'\!\approx\!\frac{A_{\rm g}}{5}\omega^5$. Coincidentally, up to the accuracy of our measurement, we find the same value for the prefactor $A_{\rm g}\!\approx\!6\!\times\!10^{-3}$ in both models. 

\begin{figure}[ht!]
  \includegraphics[width = 0.5\textwidth]{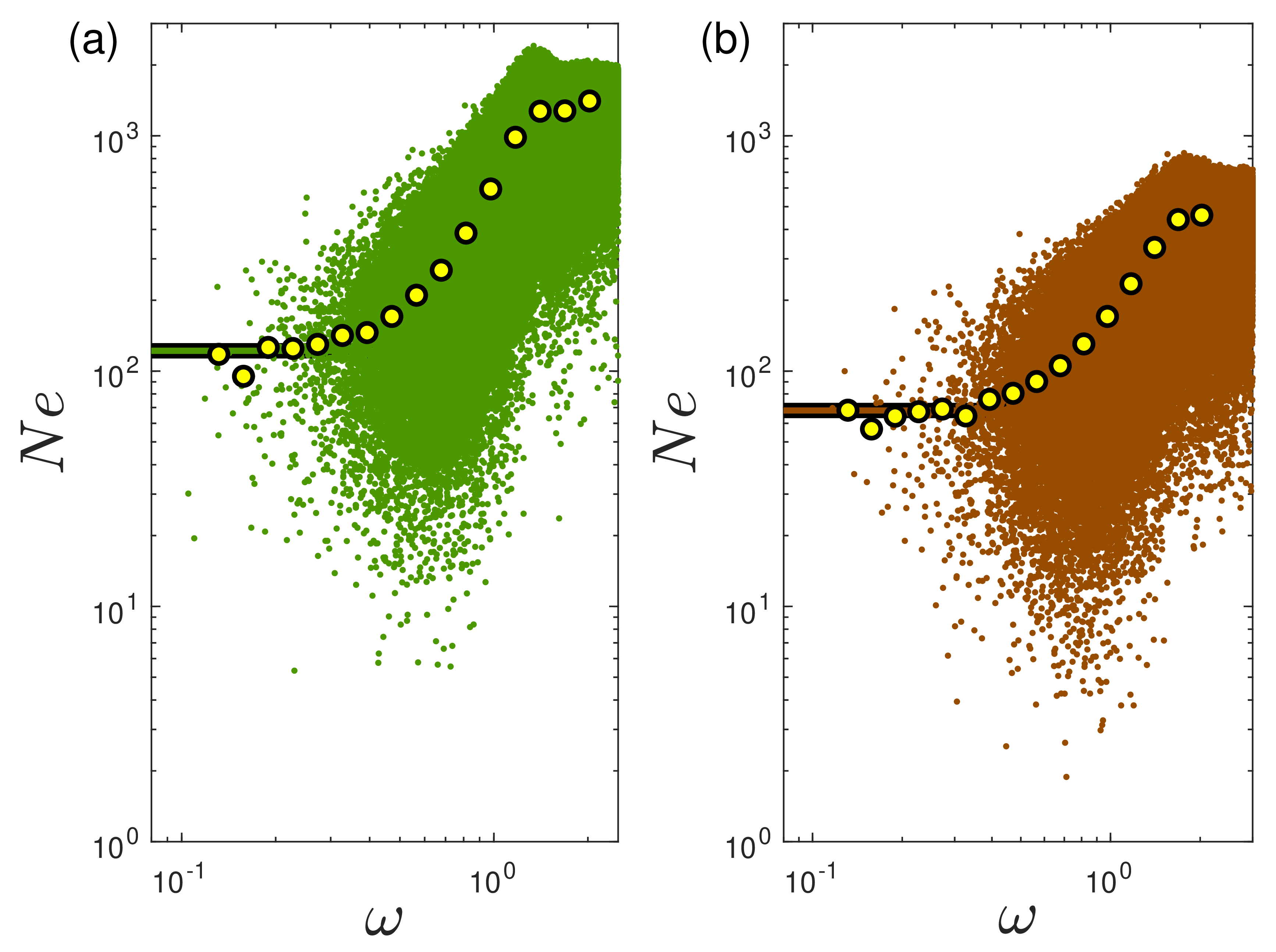}
  \caption{\footnotesize The factored participation ratio $Ne$~\cite{footnote} of vibrational modes calculated in (a) the molecular glass and in (b) the atomistic glass, scatter-plotted against those modes' frequencies $\omega$. The yellow circles represent running averages, and the horizontal lines are guides to the eye of the low-frequency plateau, corresponding to (a) $Ne_{\rm mol}\!\approx\!120$ and (b) $Ne_{\rm atom}\!\approx\!70$, and see text for further discussion.}
  \label{fig:participation_fig}
\end{figure}

What spatial properties do the modes occupying the low-frequency tails of the nonphononic VDoS have? To address this question, we scatter-plot in Fig.~\ref{fig:participation_fig} the participation ratio $e$ of vibrational modes, factored by the number of atoms $N$, against those modes' frequencies $\omega$. The products $Ne$ represent the spatial extent of these quasilocalized vibrations' core, in units of atom numbers. The yellow circles represent running averages, and the horizontal lines are guides to the eye of the low-frequency plateau that the running averages assume. This presentation clearly exposes the frequency $\tilde{\omega}_{\rm min}$ of the lowest-frequency phononic modes ($\tilde{\omega}_{\rm min}\!=\!2\pi c_{\rm s}/L\!\approx\!1.6$ in the molecular glass, and $\tilde{\omega}_{\rm min}\!\approx\!2.0$ in the atomistic glass), seen as a peak at $Ne\!\sim\!{\cal O}(N)$ in the scatter plot of the scaled participation vs.~frequency. The lowest values of $Ne$ are $\approx\!5$ in the molecular glass, much larger than the smallest values in the atomistic glass, closer to $\approx\!2$.

The low-frequency plateau values of $Ne$ are estimated as $Ne_{\rm mol}\!\approx\!120$ for the molecular glass, and $Ne_{\rm atom}\!\approx\!70$ for the atomistic glass. We note that $Ne_{\rm mol}/Ne_{\rm atom}\!\approx\!1.7\!\approx\!\chi_{\rm mol}^2/\chi_{\rm atom}^2$ ($\approx\!1.6$, cf.~Sect.~\ref{sec:dispersion}), establishing an interesting connection between macro- and microscopic mechanics, and validating that $\chi^2$ indeed represents a mesoscopic correlation volume. 

We next turn to assessing the characteristic frequency $\omega_{\rm g}$ of quasilocalized vibrations in our two glass models. To this aim we adopt the dipole-response approach as discussed above and in Refs.~\cite{cge_paper,pinching_pnas}. The approach amounts to defining a dipolar force on a pair of particles $i,j$, of the form
\begin{equation}
    \dv^{(ij)} \equiv \frac{\partial r_{ij}}{\partial\xv}\,,
\end{equation}
where $r_{ij}$ is the pairwise separation between atoms $i$ and $j$, and $\xv$ denotes the $N\dbar$-dimensional vector of particle coordinates. The linear response $\uv^{(ij)}$ to the dipolar force $\dv^{(ij)}$ reads
\begin{equation}
    \uv^{(ij)} = \calBold{H}^{-1}\cdot\dv^{(ij)}\,,
\end{equation}
where $\calBold{H}\!\equiv\!\frac{\partial^2U}{\partial\xv\partial\xv}$ is the Hessian matrix of the potential energy $U(\xv)$. Previous work~\cite{cge_paper} has shown that the response fields $\uv$ bear many structural similarities to quasilocalized vibrations. The stiffness $\kappa_{ij}$ associated with the response $\uv^{(ij)}$ reads
\begin{equation}
    \kappa_{ij} \equiv \frac{\uv^{(ij)}\cdot\calBold{H}\cdot\uv^{(ij)}}{\uv^{(ij)}\cdot\uv^{(ij)}}=\frac{\dv^{(ij)}\cdot\calBold{H}^{-1}\cdot\dv^{(ij)}}{\dv^{(ij)}\cdot\calBold{H}^{-2}\cdot\dv^{(ij)}}\,.
\end{equation}

To determine the characteristic frequency scale $\omega_{\rm g}$, we average $\kappa_{ij}$ over many pairs $i,j$ of atoms that are close-by -- but not interacting, and see discussion in Ref.~\cite{cge2_jcp2020} regarding this choice. The square root of this average defines the characteristic frequency scale $\omega_{\rm g}$, which is reported for the two glass models in Table~\ref{table:qlv} below. We find that $\omega_{\rm g}$ is lower by roughly 10\% in our molecular glass compared to its counterpart in the atomistic glass. This is surprising since the shear modulus of the molecular glass is higher compared to the shear modulus of the atomistic glass, leading to the naive and wrong expectation that $\omega_{\rm g}$ would also be higher in the molecular glass. 

With $\omega_{\rm g}$ at hand, we gain access to two additional observables related to quasilocalized vibrations. First, in Ref.~\cite{pinching_pnas} it was established that the linear size of the core of quasilocalized vibrations is given by a mesoscopic length $\xi_{\rm g}$ defined as
\begin{equation}
    \xi_{\rm g} \equiv \frac{2\pi c_{\rm s}}{\omega_{\rm g}}\,.
\end{equation}
This length bears similarity with the `boson peak' length $\xi_{\rm bp}\!\equiv\!c_{\rm s}/\omega_{\rm bp}$ where $\omega_{\rm bp}$ is the boson peak frequency~\cite{sokolov_boson_peak_scale}. Interestingly, we find that $\xi_{{\rm g,mol}}/\xi_{{\rm g,atom}}\!\approx\!\xi_{{\rm s,mol}}/\xi_{{\rm s,atom}}$ (compare 1.23 to 1.17), suggesting that the core-size $\xi_{\rm g}$ of QLVs is proportional to the dispersion length $\xi_{\rm s}$, a question to be addressed in future work.

Second, with the characteristic frequency $\omega_{\rm g}$ at hand, we are able to extract the number ${\cal N}$ per degree-of-freedom of quasilocalized modes via ${\cal N}\!=\!A_{\rm g}\omega_{\rm g}^5$, as discussed above. We find that, in the molecular glass, ${\cal N}$ is smaller by approximately 40\% compared to the atomistic glass, which --- on the face of it --- appears to be at odds with the consistently higher degree of mechanical disorder as reflected in the lengths $\xi_{\rm g}$ and $\xi_{\rm s}$, in the size $Ne$ of soft, quasilocalized vibrations, and in the mechanical disorder quantifier $\chi$. We offer a rationalization of this interesting difference in Sect.~\ref{sec:discussion}. Our results for this Section are summarized in Table~\ref{table:qlv}.

\begin{table}[h]
    \centering
    \caption{Quasilocalized vibrations' properties}
    \label{table:qlv}
    \begin{tabular}{| c | c c c c c |}
        \hline
        model & $A_{\rm g}$ & $N\langle e \rangle$ & $\omega_{\rm g}$ & $\xi_{\rm g}$ &   ${\cal N}$   \\ \hline
        molecular & $6\!\times\!10^{-3}$ & 120 & 2.85 & 7.92 & 1.1  \\ \hline
        atomistic & $6\!\times\!10^{-3}$ & 70 & 3.15 & 6.45 & 1.85  \\ \hline
    \end{tabular}
\end{table}

\section{P\lowercase{roperties of nonlinear plastic modes}}
\label{sec:nonlinear_modes}
\vspace{-0.2cm}

\subsection{Introduction and definition}
\vspace{-0.2cm}

When a glass is subject to external deformation, it undergoes plastic instabilities in which the local minimum on the potential energy landscape (PEL) develops an unstable direction in configuration space~\cite{Malandro_Lacks}. Soft, quasilocalized vibrational modes are known to be sensitive to such instabilities~\cite{lemaitre2004}, and to anticipate the loci of irreversible rearrangements in both atomistic~\cite{manning2011} and polymeric~\cite{Smessaert2014,Smessaert2015} glasses; side-by-side simulations of polymer and binary Lennard-Jones glasses have further shown that their yielding behavior shares many common features~\cite{Rottler2003}. However, a more robust characterization of plastic instabilities is provided by the nonlinear plastic modes (NPMs) framework~\cite{micromechanics2016}, which accounts for the relevant anharmonicities of the PEL. NPMs --- also termed `cubic modes'~\cite{episode_1_2020} --- are normalized displacement fields $\piv$ (referred to in what follows as `modes', not to be confused with vibrational modes) that solve the following nonlinear algebraic equation~\cite{SciPost2016,episode_1_2020}
\begin{equation}\label{eq:npms}
    \calBold{H}\cdot\piv = \frac{\calBold{H}:\piv\piv}{\calBold{U}''':\!\cdot\piv\piv\piv}\,\calBold{U}''':\piv\piv\,,
\end{equation}
where $\calBold{U}'''\!\equiv\!\frac{\partial^3U}{\partial\xv\partial\xv\partial\xv}$ is the tensor of third-order derivatives of the potential energy, $:$ denotes a double-contraction and $:\!\cdot$ a triple contraction. Solutions $\piv$ to Eq.~(\ref{eq:npms}) are obtained by minimizing the `barrier function'~\cite{micromechanics2016}
\begin{equation}\label{eq:barrier_function}
    \Delta(\zv) \equiv \frac{\left(\frac{\partial^2U}{\partial\xv\partial\xv}:\zv\zv\right)^3}{\left(\frac{\partial^3U}{\partial\xv\partial\xv\partial\xv}:\!\cdot\zv\zv\zv\right)^2}\,,
\end{equation}
which sets the scale of the potential energy barrier between adjacent local minima on the potential energy landscape~\cite{makoto_barriers_arXiv}. In other words, Eq.~(\ref{eq:npms}) is obtained via 
\begin{eqnarray}
    \frac{\partial \Delta}{\partial\zv}\bigg|_{\zv=\piv} = \zerovector\,.
\end{eqnarray}

\begin{figure}[ht!]
  \includegraphics[width = 0.5\textwidth]{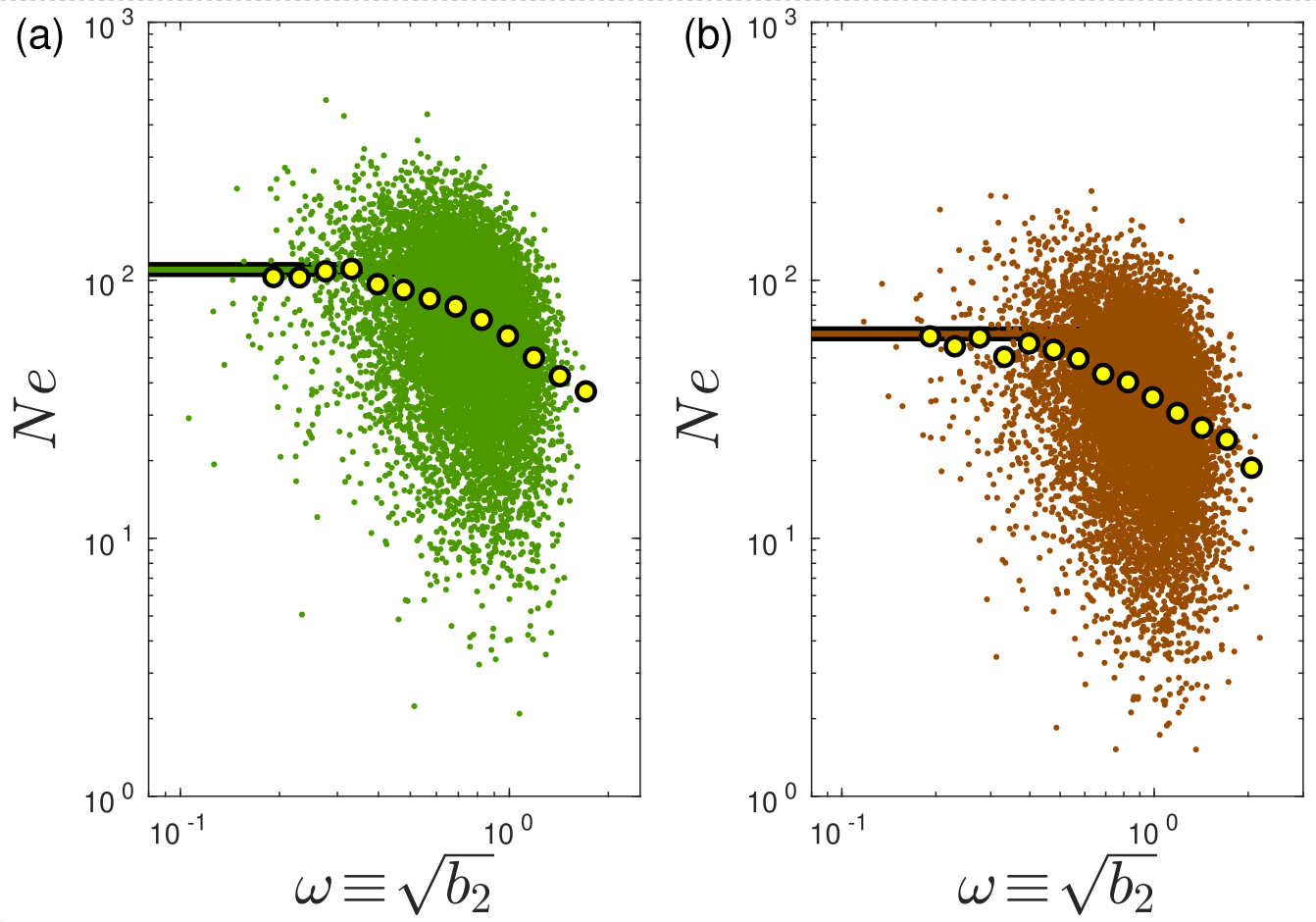}
  \caption{\footnotesize The factored participation ratio $Ne$~\cite{footnote} of nonlinear plastic modes (NPMs) calculated in (a) the molecular glass and in (b) the atomistic glass, scatter-plotted against those modes' effective frequencies $\omega\!\equiv\!\sqrt{b_2}$. The yellow circles represent running averages, and the horizontal lines are guides to the eye of the low-frequency plateau, corresponding to (a) $Ne_{\rm mol}\!\approx\!110$ and (b) $Ne_{\rm atom}\!\approx\!60$, see text for further discussion.}
  \label{fig:npm_localization_fig}
\end{figure}

\subsection{Localization properties}
\vspace{-0.2cm}

We first study the localization properties of NPMs of both the molecular and atomistic glasses. To this aim, and for subsequent analyses in this Section, we find a single, low-energy NPM for each glass sample in our ensembles, for both glass models (i.e.~a total of 10,000 NPMs per model), following the scheme described in Appendix~\ref{sec:NPM_appendix}. In Fig.~\ref{fig:npm_localization_fig} we scatter-plot the scaled participation ratio $Ne$ vs.~NPMs' effective frequencies $\omega\!\equiv\!\sqrt{b_2}$, where $b_2\!\equiv\!\calBold{H}\!:\!\piv\piv$ is the stiffness associated with the NPM $\piv$. The yellow circles are running averages, while the horizontal lines are guides to the eye for the low-frequency plateau. Similar to low-frequency QLVs, we find the low-frequency plateaus at $Ne\!\approx\!110$ for the molecular NPMs, and $Ne\!\approx\!60$ for the atomistic NPMs. In contrast to the vibrational-mode data (cf.~Fig.~\ref{fig:participation_fig}), these data show that NPMs become more localized as they stiffen, in both our molecular and atomistic glasses, whereas vibrational modes progressively hybridize with phononic modes and hence do not show the same trend.

\subsection{Coupling to external deformation}
\vspace{-0.2cm}

Close to plastic instabilities, the stiffness $b_2\!\equiv\!\calBold{H}:\piv\piv$ associated with NPMs follows an approximate equation of motion with external deformation $\bm{\epsilon}$ as~\cite{micromechanics2016}
\begin{equation}
    \frac{db_2}{d\bm{\epsilon}} \simeq -\frac{b_3}{b_2}\calBold{F}\,,
\end{equation}
where $b_3\!\equiv\!\calBold{U}'''\!:\!\cdot\,\piv\piv\piv$ quantifies the strength of cubic anharmonicity, ${\bm\epsilon}$ is the strain tensor, and 
\begin{equation}
    \calBold{F} = \piv\cdot\frac{\partial^2U}{\partial\xv\partial{\bm\epsilon}}
\end{equation}
is a tensor that quantifies the coupling of the NPM $\piv$ to an external deformation characterized by the strain tensor $\bm\epsilon$.

In Ref.~\cite{sticky_spheres1_karina_pre2021} the properties of the tensor $\calBold{F}$ were investigated in a broad variety of atomistic computer glasses, and see a related analysis in Ref.~\cite{elementary_processes_prr_2024}. Here we adopt a similar framework to compare between NPMs in our molecular and atomistic glasses. More specifically, we decompose $\calBold{F}\!=\!\calBold{F}^{\rm dil} + \calBold{F}^{\rm dev}$ into dilatational and deviatoric parts, where $\calBold{F}^{\rm dil}\!\equiv\!\calBold{I}\mbox{Tr}(\calBold{F})/\dbar$ ($\calBold{I}$ is the identity tensor) and $\calBold{F}^{\rm dev}\!\equiv\!\calBold{F}\!-\!\calBold{F}^{\rm dil}$. With this framework, we finally define the dilation-to-shear ratio ${\cal R}_{\rm ds}$ of NPMs as~\cite{elementary_processes_prr_2024}
\begin{equation}\label{eq:R_ds}
{\cal R}_{\rm ds} \equiv \frac{|\mbox{Tr}(\calBold{F}^{\rm dil})/\dbar|}{\sqrt{\frac{\dbar}{\dbar-1}\calBold{F}^{\rm dev}\!:\!\calBold{F}^{\rm dev}}}    \,,
\end{equation}
where $\calBold{F}^{\rm dev}\!\!:\!\!\calBold{F}^{\rm dev}\!\equiv\!\sum_{i,j}{\cal F}^{\rm dev}_{ij}{\cal F}^{\rm dev}_{ij}$, and notice that $\mbox{Tr}(\calBold{F}^{\rm dil})\!=\!\mbox{Tr}(\calBold{F})$. The factor $\frac{\dbar}{\dbar-1}$ corrects for the loss of one degree of freedom incurred when subtracting the trace-derived dilatational part from $\calBold{F}$, yielding an unbiased measure of the deviatoric norm per remaining degree of freedom. Finally, since $\mbox{Tr}(\calBold{F}^{\rm dil})/\dbar$ is a signed quantity whose sign is arbitrary (as $\pi$ is defined up to a sign, see Eq.~(\ref{eq:npms})), we take its absolute magnitude in the definition of ${\cal R}_{\rm ds}$, then ${\cal R}_{\rm ds}\!\ge\!0$ by construction.

\begin{figure}[ht!]
  \includegraphics[width = 0.5\textwidth]{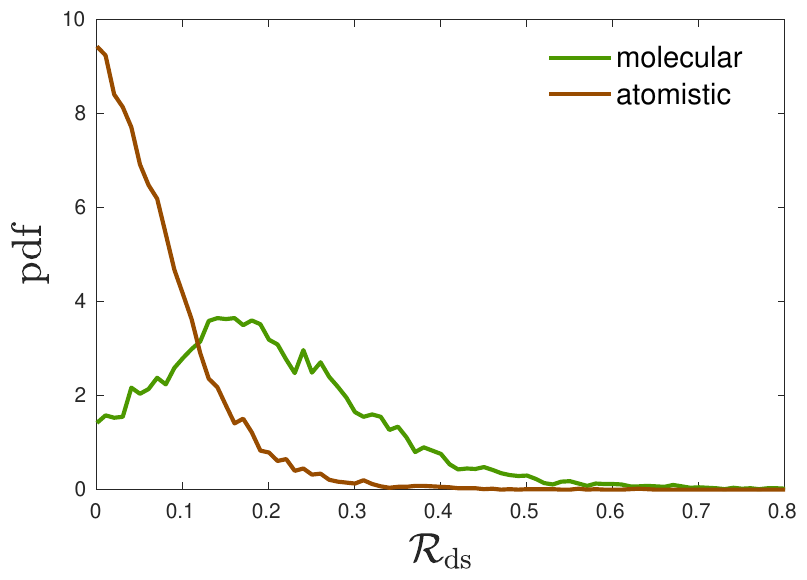}
  \caption{\footnotesize Probability distribution functions of the dilation-to-shear ratio ${\cal R}_{\rm ds}$ (cf.~Eq.~(\ref{eq:R_ds})) of nonlinear plastic modes calculated in the molecular and atomistic glasses as described in the figure legend, see text for discussion.}
  \label{fig:R_ds}
\end{figure}

In Fig.~\ref{fig:R_ds} we present the distributions $p({\cal R}_{\rm ds})$ of ${\cal R}_{\rm ds}$ calculated on our NPMs in the molecular and atomistic glasses, as described by the figure legend. Interestingly, the distributions have qualitatively different shapes, with the molecular one featuring a pronounced maximum at ${\cal R}_{\rm ds}\!\approx\!0.17$, while the atomistic distribution peaks at ${\cal R}_{\rm ds}\!=\!0$. Additionally, we find that the average values of ${\cal R}_{\rm ds}$  are significantly different: we find $\langle{\cal R}_{\rm ds}\rangle\!\approx\!0.209$ for the molecular glass, while $\langle{\cal R}_{\rm ds}\rangle\!\approx\!0.076$ in the atomistic glass, namely it is larger by a factor of $\approx\!2.7$ in the molecular glass compared to the atomistic glass. We note that a dilatational component of elementary plastic events has long been invoked in phenomenological descriptions of glassy plasticity --- in the free-volume picture~\cite{spaepen_1977}, in Argon's shear-transformation-zone theory~\cite{argon_st}, and in atomistic simulations of metallic glasses~\cite{Schuh2003} --- and has recently been characterized at the level of individual plastic modes in atomistic computer glasses~\cite{elementary_processes_prr_2024}; our results suggest that stiff molecular constraints may substantially enhance this component. This interesting difference is further discussed in Sect.~\ref{sec:discussion}.

\subsection{Planarity}
\vspace{-0.2cm}

The ratio
${\cal R}_{\rm ds}$ quantifies the shear vs.~dilation deformation modes that characterize NPMs. In addition, we further exploit our tensorial framework to characterize the degree of planarity of NPMs. This is carried out as follows: consider the eigenvalues of $\calBold{F}^{\rm dev}$, each representing the extension or compression along their associated eigenvector; order the eigenvalues of $\calBold{F}^{\rm dev}$ by magnitude, and denote the ordered set as $\lambda_1,\lambda_2,\lambda_3$ in descending order. Since $\calBold{F}^{\rm dev}$ is traceless, their sum must vanish identically. Therefore, a planar NPM is expected to have $\lambda_1\!\approx\!-\lambda_2$, and $\lambda_3\!\approx\!0$. On the other hand, a maximally non-planar NPM would feature $\lambda_2\!\approx\!\lambda_3\!\approx\!-\lambda_1/2$. We therefore define the planarity index PI as
\begin{equation}\label{eq:planarity}
    \mbox{PI} \equiv -2\lambda_2/\lambda_1 - 1\,.
\end{equation}
For a perfectly planar NPM, $\lambda_1\!\approx\!-\lambda_2$ and $\mbox{PI}\!\approx\!1$. For a completely non-planar NPM, $\lambda_1\!\approx\!-2\lambda_2$ and $\mbox{PI}\!\approx\!0$. We note that this definition differs from that of Ref.~\cite{elementary_processes_prr_2024}, there the `planarity ratio' vanishes for a perfectly planar mode.

\begin{figure}[ht!]
  \includegraphics[width = 0.5\textwidth]{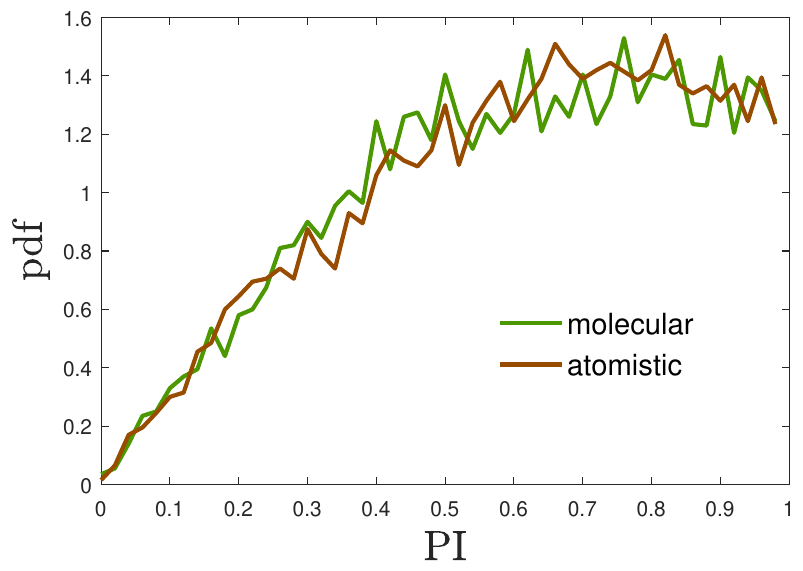}
  \caption{\footnotesize Probability distribution functions of the planarity index PI (cf.~Eq.~(\ref{eq:planarity})) of nonlinear plastic modes calculated in the molecular and atomistic glasses as described in the figure legend, see text for discussion.}
  \label{fig:PI}
\end{figure}

With these definitions at hand, we calculated the planarity index for our ensembles of NPMs of the molecular and atomistic glasses. The distributions of PIs are plotted in Fig.~\ref{fig:PI} for the molecular and atomistic glasses as described by the figure legend. We see that the distributions largely overlap, with shapes familiar from other, similar studies~\cite{elementary_processes_prr_2024}. We conclude that stiff molecular constraints do not substantially affect the degree of planarity of NPMs. We find no correlations between ${\cal R}_{\rm ds}$ and the PI of NPMs.

\subsection{Stability of local potentials}
\vspace{-0.2cm}

\begin{figure}[ht!]
  \includegraphics[width = 0.5\textwidth]{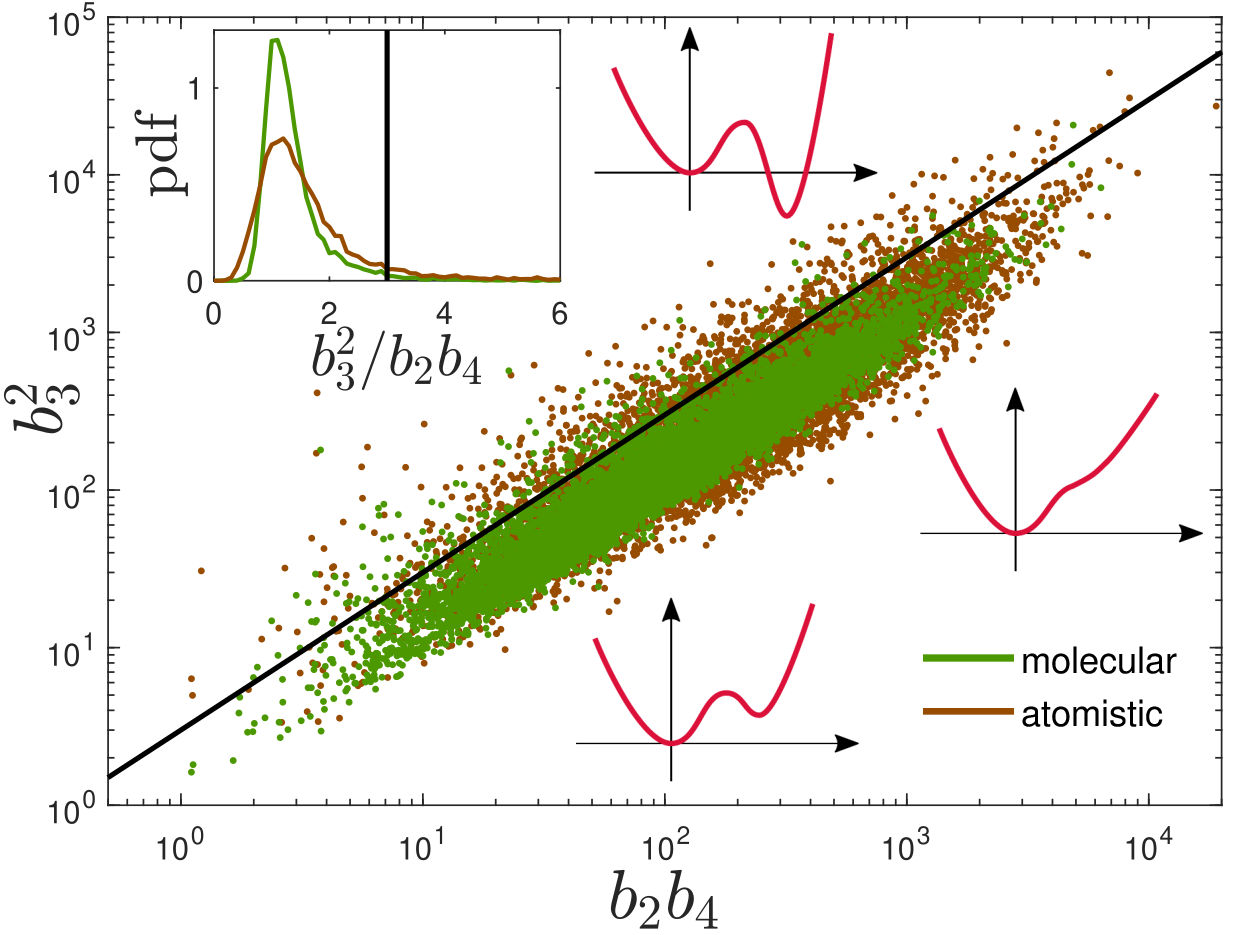}
  \caption{\footnotesize $b_3^2$ is scatter-plotted vs.~the product $b_2b_4$, see text and Eq.~(\ref{eq:local_potential}) for the definition of the coefficients $b_n$ and the local quartic potentials $\delta U_{\piv}(s)$ defined by NPMs $\piv$ of the two glass models. The cartoons illustrate the possible forms of local potentials: those with $b_3^2\!>\!3b_2b_4$ feature a second, lower-energy minimum, and are thus deemed `unstable'. The inset presents the PDFs of the dimensionless ratio $b_3^2/b_2b_4$, see text for discussion.}
  \label{fig:b2b3b4}
\end{figure}

Our last NPM analysis is concerned with the form of local potentials $\delta U_{\piv}(s)$ defined by each NPM, namely
\begin{equation}\label{eq:local_potential}
    \delta U_{\piv}(s) = \frac{1}{2}b_2 s^2 + \frac{1}{6}b_3 s^3 + \frac{1}{24}b_4 s^4 + {\cal O}(s^5)\,,
\end{equation}
where the coefficients $b_2,b_3$ were defined above, $b_4\!=\!\calBold{U''''}::\piv\piv\piv\piv$, $\calBold{U''''}\!\equiv\!\frac{\partial^4U}{\partial\xv\partial\xv\partial\xv\partial\xv}$, $::$ denotes a quadruple contraction, and we assume particles are displaced from their equilibrium configuration by $s\piv$. Given a NPM $\piv$, the coefficients $b_n$ as defined above are readily measured; they dictate the form of the local potential $\delta U_{\piv}(s)$ pertaining to each NPM. In particular, the stiffness $b_2\!>\!0$ due to the positive definiteness of the Hessian $\calBold{H}$. We also generally find $b_4\!>\!0$ for all cases studied. Importantly, if $b_3^2\!>\!3b_2b_4$, the local potential features a second minimum with lower energy~\cite{episode_1_2020,JCP_Perspective,makoto_barriers_arXiv}, as illustrated by the cartoons of local potentials in Fig.~\ref{fig:b2b3b4}; such local potentials are deemed `unstable'.  A previous study~\cite{episode_1_2020} found that glasses formed by quenching deeply supercooled states feature relatively less unstable local potentials. Therefore the fraction of unstable local potentials can be interpreted as an indicator of mechanical instability. 

In Fig.~\ref{fig:b2b3b4} we scatter plot $b_3^2$ vs.~the products $b_2b_4$ for our ensembles of NPMs of the two glass models. The two clouds largely overlap, indicating once again that our equal-footing comparison between the two models is meaningful. However, the cloud pertaining to the atomistic glass has a larger scatter, with substantially more hits above the stability line $b_3^2\!=\!3b_2b_4$. Indeed, we find that the fraction of local potentials with $b_3^2\!>\!3b_2b_4$ is approximately 8\% for the atomistic glass, whereas it is only 2.5\% for the molecular glass, namely over a factor of 3 smaller in the molecular glass. The larger scatter of the atomistic cloud is also reflected by the probability distribution functions of the dimensionless ratio $b_3^2/b_2b_4$, shown in the inset of Fig.~\ref{fig:b2b3b4} for the two glass models (the vertical line marks the stability limit $b_3^2/b_2b_4\!=\!3$); the properties of these PDFs are reported in Table~\ref{table:b2b3b4}. We conclude that, interestingly, the atomistic glass' potential energy landscape features many more unstable soft directions compared to the molecular glass, at least in terms of the quartic expansions analyzed here (cf.~Eq.~(\ref{eq:local_potential})). We do not, however, understand the origin of this difference, which is left for future investigations.

\begin{table}[h]
    \centering
    \caption{Properties of the PDFs of $b_3^2/b_2b_4$}
    \label{table:b2b3b4}
    \begin{tabular}{| c | c c c |}
        \hline
        model & mean & $\ $standard  deviation$\ $ &  unstable \%    \\ \hline
        molecular & 1.44 & 0.86 & 2.5    \\ \hline
        atomistic & 1.72 & 1.76 & 7.95    \\ \hline
    \end{tabular}
\end{table}

\section{D\lowercase{iscussion and outlook}}
\label{sec:discussion}
\vspace{-0.2cm}

In this work we set out to reveal how stiff, molecular constraints affect various properties of the vibrational modes and of the carriers of plastic-deformation -- the nonlinear plastic modes. To this aim, we employ molecular and atomistic glass forming models where the same intermolecular interaction potential of the former is also used as the interatomic interaction potential in the latter. In addition, we use the same glass-formation protocol for the two systems, and choose the number density in each system such that the hydrostatic pressure of the resulting glasses is vanishingly small. Finally, we report lengths in terms of $a_0\!\equiv\!(V/N)^{1/3}$, which, together with all other aforementioned choices, allows for a meaningful, if not perfectly controlled, comparison (cf.~Sect.~\ref{sec:model_differences}) between the mechanics of the two models. 

We find that while the shear and volumetric stiffnesses of the molecular glass are higher than in the atomistic glass, the former features a higher degree of mechanical disorder than the latter. This is consistently and systematically reflected by the higher values of $\chi,Ne,\xi_{\rm s}/a_0$ and $\xi_{\rm g}/a_0$ --- which are all dimensionless measures of mechanical disorder ---, of the molecular glass compared to those found in the atomistic glass. Interestingly, and consistent with the higher mechanical disorder of our molecular glass, it features a slightly lower mesoscopic stiffness $\omega_{\rm g}^2$, despite its higher macroscopic stiffness.

How can the higher degree of mechanical disorder of the molecular glass be rationalized? For the sake of the following argument, consider that the stiffness of intramolecular interactions is infinite, namely consider perfectly rigid molecules. Then, for $N$ atoms forming $N/3$ molecules, the number of degrees of freedom is $2N$ since every molecule contributes 6 degrees of freedom -- 3 translations and 3 rotations. This should be compared to $3N$ degrees of freedom in our atomistic model, namely a factor of 2/3 less. 

Consider now the disorder quantifier $\chi$; the factor $\sqrt{N}$ in the definition of $\chi$ (cf.~Eq.~(\ref{eq:chi})) can be interpreted as the square root of the effective number of degrees of freedom. Then, one would need to correct the measured value of our molecular glass by $\sqrt{2/3}\!\approx\!0.816$, leading to $\chi_{\rm mol}\!\approx\!5.1\!\to\!\sqrt{2/3}\!\times\!5.1\!\approx\!4.15$, not far from the value measured for our atomistic glass $\chi_{\rm atom}\!\approx\!4.0$. Following this same rationale, consider next the number density ${\cal N}$ of soft, quasilocalized excitations (cf.~Sect.~\ref{sec:qlv}). ${\cal N}$ is, by construction, a density per degree of freedom. We should therefore correct this number using the effective number of degrees of freedom, namely ${\cal N}_{\rm mol}\!\approx\!1.1\!\to\!(3/2)\!\times\!1.1\!\approx\!1.65$, not far from ${\cal N}_{\rm atom}\!\approx\!1.85$. Finally, consider the dimensionless size $Ne$ of quasilocalized vibrations (cf.~Sect.~\ref{sec:qlv}); if we substitute $N$ with the effective number of degrees of freedom, one finds $Ne_{\rm mol}\!\approx\!120\!\to\!(2/3)\!\times\!120\!\approx\!80$, not far from $Ne_{\rm atom}\!\approx\!70$.

We conclude that accounting for the effective number of degrees of freedom explains some of the observed differences in mechanical disorder between the molecular and atomistic glasses reasonably well. Once the effective number of degrees of freedom is correctly accounted for, both glasses feature very similar degrees of mechanical disorder. We emphasize that this rescaling is a leading-order correction: it treats the $2N$ low-frequency degrees of freedom of the molecular glass as statistically equivalent to the $3N$ degrees of freedom of the atomistic glass, which is a plausible but not exact assumption: it neglects the finite intramolecular stiffness of the molecular model. Future research should investigate whether counting arguments such as those presented above can explain differences in vibrational and elastic properties of glasses made of non-compact molecules such as elongated rods or star-like composite particles.

Notwithstanding the above discussion, the most striking distinction we have found between the two models lies in the strength of coupling of plastic modes to dilatational strain. The distribution of the dilation-to-shear ratio ${\cal R}_{\rm ds}$ differs qualitatively between the molecular and atomistic glasses (cf.~Fig.~\ref{fig:R_ds}), with the former featuring a pronounced peak at ${\cal R}_{\rm ds}\!\approx\!0.17$, which is totally absent from the corresponding distribution of the atomistic glass. Pressure-sensitive yielding is a long-established feature of molecular and polymeric glasses: it is built into classical theories of polymer-glass plasticity~\cite{Argon1973}, and simulations of model polymer glasses have shown that their yield surface is well described by a pressure-modified von Mises criterion, with a pressure coefficient that is nearly insensitive to the details of the model~\cite{Rottler2001}. Our ${\cal R}_{\rm ds}$ measurements offer a microscopic, mode-level counterpart to this macroscopic phenomenology, and suggest that volumetric plastic response might also differ in an interesting way between our molecular and atomistic glasses, a matter to be addressed in future work.

Another interesting question for future work could be to vary the ratio between the intramolecular bond length and the intermolecular bond length (which is approximately unity in our molecular model, cf.~Fig.~\ref{fig:potential_fig}), and to investigate the implications on the corresponding glasses' mechanics.

We emphasize finally that the term ``molecular glass'' is often used to refer to polymeric glasses, which are obviously molecular in the literal sense. Our molecular model, however, is not intended to represent that class of materials. In polymer glasses --- and, more broadly, in systems built from long, flexible molecules --- the elastic and mechanical response is strongly shaped by chain connectivity and by the vast conformational phase space explored by the chains~\cite{Hoy2007,Giuntoli2018}. The stiff, few-atom molecules employed here possess very little conformational freedom; their elasticity is predominantly energetic in origin. Our results therefore pertain to the class of glasses whose mechanics is dominated by local, energetic elasticity, and should not be extrapolated to polymeric or other conformationally rich molecular glasses without further study. Extending the present comparison toward longer and more flexible molecules, where chain-connectivity effects progressively switch on, is an interesting avenue for future work.

\section*{A\lowercase{cknowledgments}}
\vspace{-0.2cm}

We thank Eran Bouchbinder for providing useful feedback. 

\vspace{-0.2cm}

\section*{D\lowercase{ata availability}}
\vspace{-0.2cm}
The data that support the findings of this article are available from
the corresponding author upon reasonable request.

\appendix

\section{Decomposition of molecular motion}
\label{sec:molecular_motion_appendix}
\vspace{-0.2cm}

The analysis described below is applied to vibrational modes $\Psiv$ of the molecular glass model, cf.~Fig.~\ref{fig:full_vdos} in Sect.~\ref{sec:full_vdos}. Denote by $\psiv$ the components of $\Psiv$ pertaining to a single molecule. We first decompose each $\psiv$ into translational, rotational and intramolecular deformation components as 
\begin{equation}
    \psiv = \psiv_{\rm tran} + \psiv_{\rm rot} + \psiv_{\rm def}\,.\label{eq:particle_motion_decomp}
\end{equation}
The translation component for our equal-mass atoms reads
\begin{equation}
    \psiv_{\rm tran} = \frac{1}{3}\sum_\alpha \left( \sum_{i=1}^3\psiv_i\cdot\hat{\alpha}\right)\hat{\alpha}\,,
\end{equation}
where $\alpha\!=\!x,y,z$ denotes Cartesian components, the sum in the brackets runs over the 3 atoms of the molecule, and $\psiv_{\rm tran}$ is understood to be assigned to each atom in the molecule. 

We next obtain the rotational component $\bm{\psi}_{\rm rot}$ via the
Kabsch algorithm~\cite{Kabsch, kabsch1978}: we find the proper rotation $\bm{R}$
minimizing 
\begin{equation}
    \sum_{i=1}^{3}\big\|\bm{R}\cdot\bm{r}_i
    - \big(\bm{r}_i+\bm{\psi}_i-\bm{\psi}_{{\rm tran},i}\big)\big\|^2\,,
\end{equation}
i.e.\ the rotation that best maps the molecule's reference geometry onto
its translation-removed displaced geometry, where $\bm{r}_i$ is the
position of atom $i$ relative to the molecular center of mass. The
rotational displacement is given by
\begin{equation}
    \bm{\psi}_{{\rm rot},i} = \big(\bm{R}-\calBold{I}\big)\cdot\bm{r}_i\,,
\end{equation}
with $\calBold{I}$ the identity, and the deformation component is defined
as the remainder,
\begin{equation}
    \bm{\psi}_{\rm def} = \bm{\psi}-\bm{\psi}_{\rm tran}-\bm{\psi}_{\rm rot}\,.
\end{equation}

Because vibrational displacements are infinitesimal, the Kabsch rotation $\bm{R}$ reduces to
an infinitesimal rotation, and the three components
$\bm{\psi}_{\rm tran}$, $\bm{\psi}_{\rm rot}$ and $\bm{\psi}_{\rm def}$
become mutually orthogonal, allowing us to decompose
\begin{equation}\label{eq:motion_components}
    S \equiv ||\psiv||^2 =  S_{\rm tran} + S_{\rm rot} + S_{\rm def}\,,
\end{equation}
without cross terms, where $S_{\rm tran}\!\equiv\!||\psiv_{\rm tran}||^2$, and likewise for the rotational and deformational components. The percentages reported in Fig.~\ref{fig:full_vdos} are defined as
\begin{equation}
    P_{\rm tran} \equiv 100\times\frac{S_{\rm tran}}{S}\,,
\end{equation}
and likewise for rotational and deformational components. These percentages are then averaged over all molecules in the glass.

\vspace{0.2cm}

\section{Obtaining NPMs}
\label{sec:NPM_appendix}
\vspace{-0.2cm}

We follow the approach presented in Ref.~\cite{episode_1_2020} to obtain low-energy NPMs. Finding solutions to Eq.~(\ref{eq:npms}) is most readily done by minimization of the cost function~\cite{micromechanics2016,episode_1_2020,makoto_barriers_arXiv}
\begin{equation}
    \Delta(\zv) = b_2^3(\zv)/b_3^2(\zv)\,,
\end{equation}
where
\begin{eqnarray}
b_2(\zv) & = & \frac{\partial^2U}{\partial\xv\partial\xv}:\zv\zv\,, \nonumber \\
b_3(\zv) & = & \frac{\partial^3U}{\partial\xv\partial\xv\partial\xv}:\!\cdot\,\zv\zv\zv\,, \nonumber 
\end{eqnarray}
$:$ denotes a double-contraction, and $:\!\cdot$ a triple contraction. The minimization of $\Delta(\zv)$ requires choosing a meaningful initial condition $\zv_0$; we follow \cite{episode_1_2020} and minimize $\Delta(\zv)$ starting from 6 different $\zv_0$'s, of the form
\begin{equation}
    \zv_0 = \calBold{H}^{-1}\cdot\frac{\partial^2U}{\partial \xv\partial{\bm\epsilon}}\,,
\end{equation}
where we set $\bm\epsilon$ to represent simple and pure shear in the XY,YZ and XZ planes. Each such choice yields a NPM $\piv$; we select the softest one, i.e.~the one with the smallest associated stiffness $b_2(\piv)$.


\begin{thebibliography}{77}
\bibitem [{foo({\natexlab{b}})}]{footnote}%
  \BibitemOpen
  \bibinfo {note} {The participation ratio $e(\Psiv)$ of a normalized mode
  $\Psiv$ is conventionally defined as
  $e(\Psiv)\!\equiv\![N\sum_i(\Psiv_i\!\cdot\!\Psiv_i)^2]^{-1}$.}\BibitemShut
  {Stop}%
\end{thebibliography}

%

\end{document}